\documentclass[10pt]{article}

\usepackage{amsmath}
\usepackage{amssymb}
\usepackage{graphicx}

\usepackage{authblk}
\usepackage{geometry}
\newlength{\bibitemsep}
\newlength{\bibparskip}
\let\oldthebibliography\thebibliography
\renewcommand\thebibliography[1]{%
  \oldthebibliography{#1}%
  \setlength{\parskip}{\bibitemsep}%
  \setlength{\itemsep}{\bibparskip}%
}

\title{Errors in PDH offset locking due to\\spurious spectral features}

\author[1,2,a,*]{Roame Hildebrand}
\author[1,2,a]{Wance Wang}
\author[1,2]{Connor Goham}
\author[3]{Alessandro Restelli}
\author[2,4,5]{Joseph W. Britton}

\affil[1]{Institute for Research in Electronics and Applied Physics, University of Maryland, College Park, MD 20742, USA}
\affil[2]{Department of Physics, University of Maryland, College Park, MD 20742, USA}
\affil[3]{Joint Quantum Institute, University of Maryland and the National Institute of Standards and Technology, College Park, MD, USA}
\affil[4]{DEVCOM Army Research Laboratory, Adelphi, MD, USA}
\affil[5]{Quantum Technology Center, University of Maryland, College Park, MD, USA}
\affil[a]{These authors contributed equally to this work.}
\affil[*]{Email: roame@umd.edu}

\date{}

\begin{document}
\maketitle

\begin{abstract}
The Pound-Drever-Hall (PDH) technique is widely used to stabilize the frequency of lasers. Here we report on a routinely underestimated source of error in PDH offset-locking: a shift in the lock point due to the unintended interaction between residual optical sidebands and higher-order spatial modes in misaligned Fabry-P\'erot cavities. Significant frequency deviations---up to $50\%$ of the cavity linewidth---can arise when the optical offset is obtained from a sinusoidally driven EOM. We measure this deviation experimentally, find agreement with a simple model, and show how a spectrally-pure frequency offset can reduce the deviation by an order of magnitude. Our findings draw attention to a systematic effect of importance to precision optical spectroscopy, optical clocks, and quantum information science.
\end{abstract}

\section{Introduction}

The PDH technique is a cornerstone of modern precision optics \cite{drever_laser_1983,black_introduction_2001}. It enables laser stabilization for optical atomic clocks \cite{ludlow_optical_2015}, precision spectroscopy \cite{rabga_implementing_2023}, quantum information science \cite{benhelm_ca_2009,allcock_omg_2021}, and gravitational wave detection \cite{tse_quantum-enhanced_2019}. Errors due to residual amplitude modulation (RAM), etalons and thermally-driven cavity drift have been extensively studied and a variety of mitigation strategies are routinely deployed \cite{kedar_synthetic_2024,li_reduction_2016,boyd_basic_2024,zeyen_pound-drever-hall_2022}. A key extension of the technique is PDH offset locking \cite{thorpe_laser_2008,rabga_implementing_2023}, which allows a laser to be stabilized at a controlled offset from a cavity resonance. Offset locking introduces additional complexity in the form of optical spectral components---residual sidebands---that can interact with the cavity in unintended ways. 

It is well known that deviations can arise in a PDH-locked laser due to unwanted features in the optical spectrum incident on the reference cavity. A variety of schemes were explored that simplify the probe laser spectrum in the context of offset locking \cite{thorpe_laser_2008,bai_electronic_2017,rabga_implementing_2023,tu_quadrature_2024}. However, there is no published analysis that quantifies the magnitude of frequency deviation. Here we experimentally measure the frequency deviation of a PDH offset-locked laser with realistic laser-cavity misalignment and probe light spectrum. We find excellent agreement with a simple analytic model and show how serrodyne modulation can mitigate the effect \cite{kohlhaas_robust_2012,hildebrand_spectrally-pure_2024}. 

\section{A model for realistic PDH offset locking }

We consider the locked laser's frequency deviation due to imperfect laser-cavity mode matching and unwanted spectral features in probe light. For clarity, we denote optical cavity mode frequencies with $\nu$ (in Hz) and spectral features of the laser beam probing the cavity with $\xi$ (in Hz).

Fabry-P\'erot (FP) optical cavities are widely used as ultra-stable references for laser-frequency stabilization and linewidth narrowing. Ideally, the input Gaussian laser beam would be properly mode-matched to excite only the cavity's fundamental transverse mode $\mathrm{TEM}_{00}$; however, tip-tilt and transverse displacement misalignments permit excitation of higher-order Hermite--Gaussian modes $\mathrm{HG}_{mn}$($m,n\in\{\mathbb{Z}>0\}$). Furthermore, incorrect incident beam waist size or waist axial position permit excitation of Laguerre-Gaussian modes $\mathrm{LG}_{pl}$ ($p,l\in\{\mathbb{Z}>0\}$; $p$ enumerates radial mode order, $l$ enumerates angular mode order) \cite{anderson_alignment_1984}. These higher-order modes appear on the positive side of the fundamental mode at $\nu_{00}+k\nu_{h}$, where $k=m+n$ for $\mathrm{HG}_{mn}$ mode, $k=2p+l$ for $\mathrm{LG}_{pl}$ modes and $\nu_{00}$ is a $\mathrm{TEM}_{00}$ mode frequency. The frequency $\nu_{h}$ only depends on cavity geometry

\begin{equation}
\nu_{h}=\frac{1}{\pi}\nu_{\mathrm{FSR}}\arccos\left[\sqrt{\left(1-\frac{d}{R_{1}}\right)\left(1-\frac{d}{R_{2}}\right)}\right]
\end{equation}
Here we assume a cavity of length $d$, spherical mirrors with radius of curvature $R_{1}$ \& $R_{2}$, and free spectral range (FSR) $\nu_{\mathrm{FSR}}=c/2d$. Consider the lowest-order modes that emerge in the event of misalignment: $\mathrm{HG}_{10},\mathrm{HG}_{01}$ at $+\nu_{h}$ and $\mathrm{LG}_{10}$ at $+2\nu_{h}$. These higher-order resonances can be observed on the transmitted (or reflected) light when scanning the probe laser frequency. Let $C_{k}=V_{00}/V_{k}$ be the alignment contrast where $V_{00}$ is the photodetector (PD) signal for the $\mathrm{TEM}_{00}$ mode and $V_{k}$ is the PD signal for the mode offset by $k\nu_{h}$. 

\begin{figure}
\begin{centering}
\includegraphics[width=0.65\columnwidth]{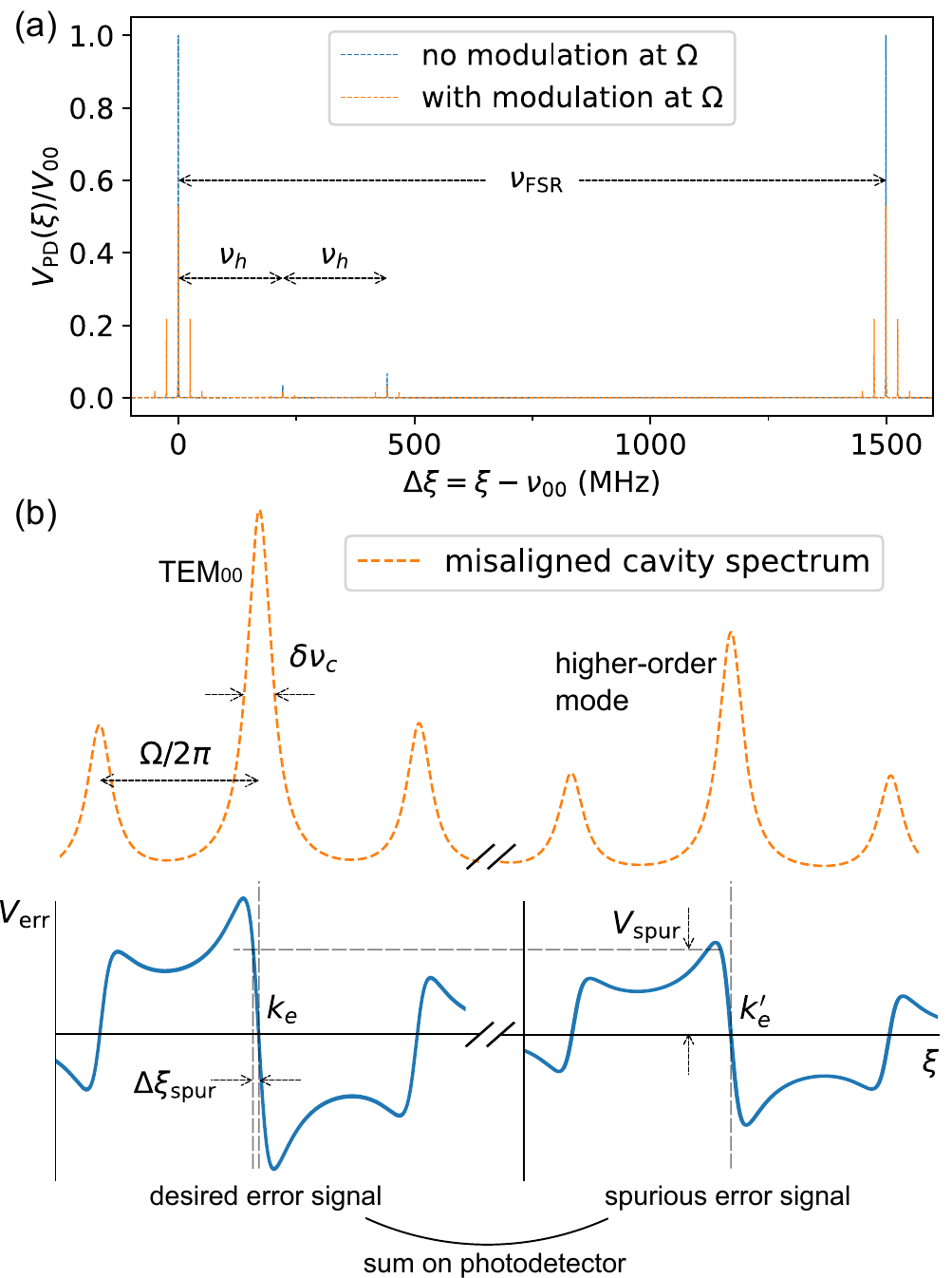}
\par\end{centering}
\centering{}\caption{\protect\label{fig:misaligned-spectrum}This figure illustrates the frequency deviation in a PDH-locked laser due to the interplay between imperfect cavity mode matching and unwanted spectral features in the offset laser spectrum. (a) Illustration of PD signal for light transmitted by a misaligned cavity (Eq.$\text{ }$\ref{eq:misaligned-fpi-spectrum}). The horizontal axis is the incident-laser detuning relative to a $\text{TEM}_{00}$ cavity mode at $\nu_{00}$. Here we assume a cavity with $d=100\,\mathrm{mm}$, $R_{1}=500\,\mathrm{mm}$, $R_{2}=\infty$, $\delta\nu_{c}=113.4\,\mathrm{kHz}$, $C_{1}=30$, $C_{2}=15$ and $C_{k}=\infty$ for $k>2$. The orange (blue) lines show the transmitted PD spectrum with (without) PDH phase modulation at $\Omega/2\pi=25\,\mathrm{MHz}$ and depth $\beta=1.082\text{ rad}$ where the PDH error signal slope $k_{e}$ (V/Hz) is maximized. (b) Illustration of the mechanism underlying a shift in the PDH locking point. The left-hand (right-hand) side shows a zoom-in around $\Delta\xi=0$ ($\Delta\xi=\nu_{h}$); the upper orange trace is the PD signal and the lower blue trace is the resulting PDH error signal with slope $k_{e}$ ($k_{e}^{\prime}$). Ideally, the laser is locked to the left-hand feature so that $V_{\text{err}}=0$ and the right-hand feature is absent. The dashed horizontal line marks the spurious PDH error signal due to laser light near the higher-order mode at $\nu_{h}$ which gives rise to an offset $V_{\mathrm{spur}}$ in the PDH error signal and a corresponding spurious lock point shift $\Delta\xi_{\mathrm{spur}}=V_{\mathrm{spur}}/k_{e}$. Note that $\delta\nu_{c}$, $\Omega/2\pi$, and the contrasts for this figure are exaggerated to articulate certain effects. Physically realistic parameters are used in the calculation in Figure \ref{fig:lock-shift-calc-plot}}
\end{figure}

If a probe laser of frequency $\xi$ is scanned across the cavity FSR, the transmitted PD voltage signal has the spectrum
\begin{align}
V_{\text{PD}}(\xi) & =V_{00}\sum_{n}L(\xi,n\nu_{\mathrm{FSR}})\nonumber \\
& \quad+\frac{1}{C_{1}}L(\xi,n\nu_{\mathrm{FSR}}+\nu_{h})\label{eq:misaligned-fpi-spectrum}\\
& \quad+\frac{1}{C_{2}}L(\xi,n\nu_{\mathrm{FSR}}+2\nu_{h})\nonumber 
\end{align}
where
\[ L(x,x^{\prime})=\frac{1}{1+\left(\frac{x-x^{\prime}}{\delta\nu_{c}/2}\right)^{2}} \]
is a Lorentzian lineshape with cavity FWHM linewidth $\delta\nu_{c}$. As an example, consider $d=100\,\mathrm{mm}$, $R_{1}=500\,\mathrm{mm}$ and $R_{2}=\infty$ so that $\nu_{\mathrm{FSR}}=1.5\,\mathrm{GHz}$ and $\nu_{h}=0.221\,\mathrm{GHz}$. We observe that when $C_{1}\sim30$ and $C_{2}\sim15$ (typical for our $870\text{ nm}$ laser-cavity system), laser-cavity alignment is stable for more than a week at a time. Better alignment is possible but requires more frequent adjustment. In practice, transverse-position and angular-tilt misalignment are easier to reduce, while beam waist size and waist axial position are harder; typically $C_{1}>C_{2}$. Figure$\,$\ref{fig:misaligned-spectrum}(a) shows an example of a misaligned cavity spectrum.

Here we show how an undesired spectral feature at frequency $\xi^{\prime}$ and power $P^{\prime}$ in the offset laser light can cause a spurious frequency shift in a PDH-locked laser when the laser is misaligned to the reference cavity (Fig.$\text{ }$\ref{fig:misaligned-spectrum}(b)). Let $\Delta\xi$ denote the difference in frequency between the shifted laser frequency and the target locking feature of the cavity. For an idealized PDH lock, a small frequency deviation $\Delta\xi\ll\delta\nu_{c}/2$ in the offset laser beam with power $P_{00}$, generates a PDH error signal with zero time average\cite{black_introduction_2001} $\left<V_{\text{err}}\right>=0$; that is, the laser is locked. If $\xi^{\prime}$ overlaps with a higher-order cavity mode near $\nu_{00}+k\nu_{h}$, we get a spurious error signal voltage $V_{\mathrm{spur}}$ (see Fig.$\text{ }$\ref{fig:misaligned-spectrum}). If the spur is at $\nu_{00}+k\nu_{h}-\delta\nu_{c}/2$ the error is maximal 
\begin{equation}
V_{\mathrm{spur}}^{\text{max}}=k_{e}^{\prime}\frac{\delta\nu_{c}}{2}
\end{equation}
where the error-signal slope at the higher order cavity mode is $k_{e}^{\prime}\propto P^{\prime}/C_{k}$. Since the PD output is a sum of the desired and spurious signals, the shift in the lock point due to $V_{\mathrm{spur}}^{\text{max}}$ is 
\begin{equation}
\Delta\xi_{\mathrm{spur}}^{\text{max}}=V_{\mathrm{spur}}^{\text{max}}/k_{e}=\frac{1}{C_{k}}\frac{P^{\prime}}{P_{00}}\frac{\delta\nu_{c}}{2}\label{eq:lock-shift-due-to-fake}
\end{equation}

\begin{figure*}[p]
\begin{centering}
\includegraphics[width=0.85\columnwidth]{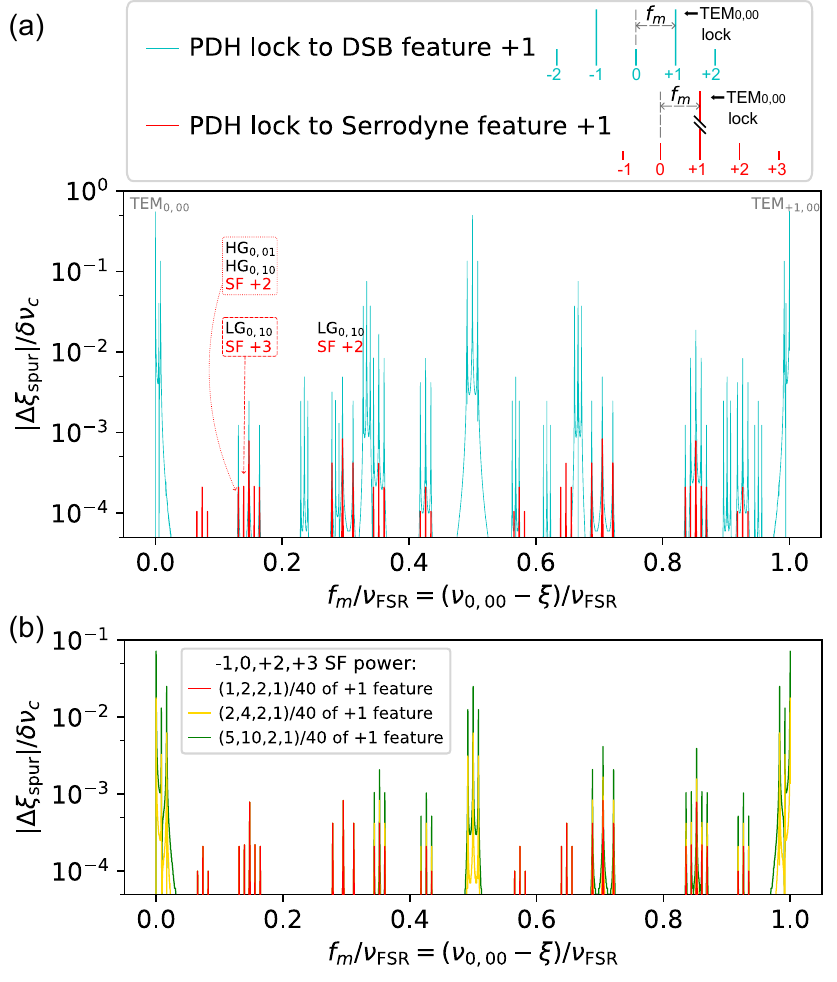}
\par\end{centering}
\caption{\protect\label{fig:lock-shift-calc-plot}Here we plot the shift in a PDH-locked laser due to interaction of the $\mathrm{HG}_{01}$, $\mathrm{HG}_{10}$ and $\mathrm{LG}_{10}$ cavity modes with unwanted spectral features of the EOM-offset laser light. Here we assume a cavity with $d=100\,\mathrm{mm}$, $R_{1}=500\,\mathrm{mm}$, $R_{2}=\infty$, $\delta\nu_{c}=113.4\,\mathrm{kHz}$, $\Omega/2\pi=25\,\mathrm{MHz}$, $C_{1}=30$, $C_{2}=15$ and $C_{k}=\infty$ for $k>2$. $f_{m}$ is DSB or serrodyne EOM modulation frequency. $f_{m}=\nu_{0,00}-\xi$ for an offset lock to a $\text{TEM}_{0,00}$ mode. The blue trace corresponds to single-tone drive of the EOM with depth $\beta\approx1.84\text{ rad}$ and PDH locking the $\text{TEM}_{0,00}$ cavity mode to the upper (+1) optical sideband. The calculation includes the $0,\pm1,\pm2$ spectral features emitted by the EOM. The orange trace corresponds to serrodyne drive of the EOM and PDH locking the $\text{TEM}_{0,00}$ cavity mode to the $+1$ feature. The calculation includes the $0,\pm1,+2$ and $+3$ serrodyne features (SF). The relative peak heights follow those observed experimentally: relative to the $+1$ feature, $0,+2$ are 13 dB lower and $-1,+3$ are $16\text{ dB}$ lower. Here we distinguish cavity modes in different FSR ranges by adding a new index $q$ (e.g., $\text{TEM}_{q,00}$) because spurious features will interact with modes spanning multiple FSR. The vertical axis shows the relative lock point shift (see Fig.$\text{ }$\ref{fig:misaligned-spectrum}(b)). As an example, $\Delta\xi_{\mathrm{spur}}$ around $f_{m}=\nu_{h}$ arises due to the interplay between SFs and cavity modes indicated in the red dashed boxes. (b) Lock point shift with symmetric (red trace) and asymmetric (yellow and green) SFs. The red trace, which is the same one in plot (a), has equal power in -1 and +2 features, as well as in 0 and +3 features. In this case $\Delta\xi_{\mathrm{spur}}$ is zero at multiples of $\nu_{\mathrm{FSR}}/2$. The other two traces are more practical examples where SFs are asymmetric.}
\end{figure*}

We are now in a position to compare generation of offset laser light by simple sinusoidal modulation and serrodyne. Following the nomenclature in \cite{thorpe_laser_2008} we refer to the former as the dual-sideband (DSB) approach. Recall that for DSB the optical power in the $\pm1$ sidebands is maximal (and optimal) for a modulation depth $\approx1.84\text{ rad}$. In the DSB case if the laser is locked to the $+1$ feature, the worst-case configuration is when the $-1$ sideband ($P^{\prime}/P_{00}=1$) overlaps with the adjacent $\text{TEM}_{00}$ mode which gives $\Delta\xi_{\mathrm{spur}}^{\text{max}}=\delta\nu_{c}/2$. The worst-case interaction with higher-order cavity modes is between the most prominent misalignment feature, say with $C_{k}=15$, and the -1 sideband, yielding $\Delta\xi_{\mathrm{spur}}^{\text{max}}=\delta\nu_{c}/30$. Contrast this with realistic performance of a serrodyne-derived offset where $P^{\prime}/P_{00}=1/10$: the corresponding shifts are $\Delta\xi_{\mathrm{spur}}^{\mathrm{max}}=\delta\nu_{c}/20$ and $\Delta\xi_{\mathrm{spur}}^{\mathrm{max}}=\delta\nu_{c}/300$ respectively. To better compare the DSB and serrodyne modulation schemes, Figure$\text{ }$\ref{fig:lock-shift-calc-plot}(a) shows a calculation of the offset-frequency-dependent laser-lock shift for realistic parameters.

\section{Frequency deviation measurement}
\begin{figure}[h]
\begin{centering}
\includegraphics[width=.6\columnwidth]{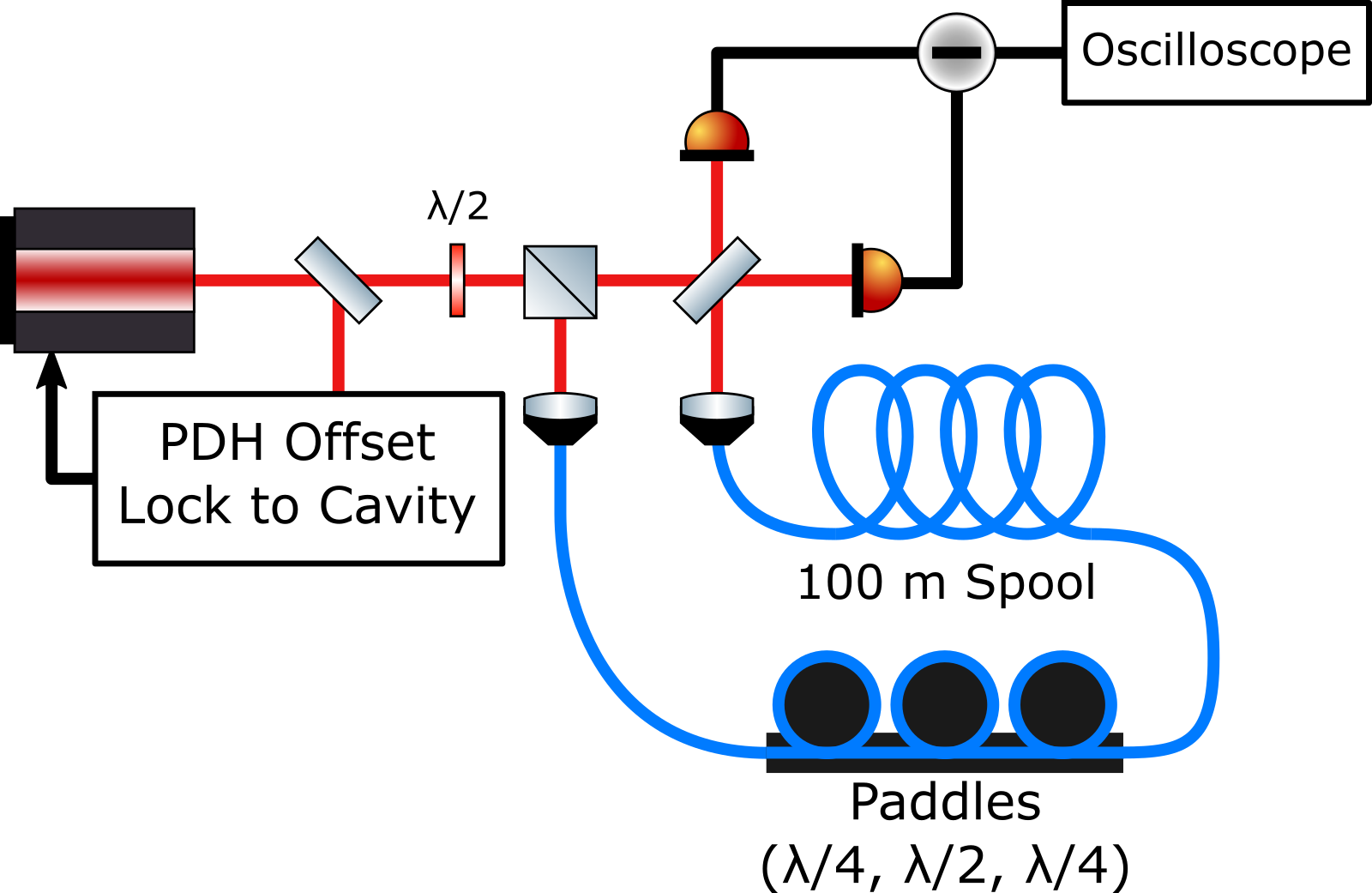}
\par\end{centering}
\caption{\protect\label{fig:unbalanced-mzi}Diagram of our OFA. Red lines represent free-space beam propagation, blue lines represent propagation in non-PM fiber. The long arm of the unbalanced MZI consists of a spool of SMF-28 fiber encapsulated in spray foam to reduce its sensitivity to vibration. Light exiting the MZI was incident on a balanced photodiode pair (Thorlabs PDB450A) and digitized with an oscilloscope.}
\end{figure}

To measure the predicted spurious frequency excursions, we need a linear optical frequency analyzer (OFA) with sensitivity at the $\sim10\text{ kHz}$ level. For this we used an unbalanced Mach-Zehnder interferometer (MZI) with optical path length difference $\Delta L$ (Fig. \ref{fig:unbalanced-mzi}). A linear frequency sweep of the laser light incident on the OFA results in a voltage out of the balanced photodetector proportional to $\cos^{2}(\Delta\phi/2)-\sin^{2}(\Delta\phi/2)=\cos(\Delta\phi)$ where $\Delta\phi=\frac{2\pi}{c}\xi_{0}\Delta L$ and $\xi_{0}$ is the unmodulated laser frequency. Any nonlinear frequency excursions during the sweep are obvious in the residual. This technique requires that $\Delta L$ remain stable for the duration of the frequency sweep.

Our test setup consists of a single-mode 822 nm ECDL (Toptica DLC Pro), which is PDH offset locked to an optical cavity (Stable Laser Systems). The PDH feedback is tuned so that the laser stays continuously locked to the cavity during a frequency sweep. We implemented a DSB frequency offset using a computer-controlled RF source that modulates an EOM. The PDH lock RF frequency was $25\text{ MHz}$. Our cavity has $R_{1}=0.5\text{ m}$, $R_{2}=\infty$. We estimated that $\nu_{FSR}=1497.0\pm0.2\text{ MHz}$ by observing the cavity's transmission spectrum while adjusting $f_{m}$ with the laser unlocked; uncertainty is taken to be roughly the linewidth of the cavity. We performed multiple cavity ring-down measurements to estimate the cavity's linewidth, yielding $\delta\nu_{c}=113.4\pm3.9\text{ kHz}$ (reported error is statistical). A fraction of the unmodulated laser light is sent to the MZI with $\Delta L=142.4\pm1.2\text{ m}$ (reported error is statistical).

\begin{figure*}[!h]
\centering
\begin{centering}
\includegraphics[width=1\columnwidth]{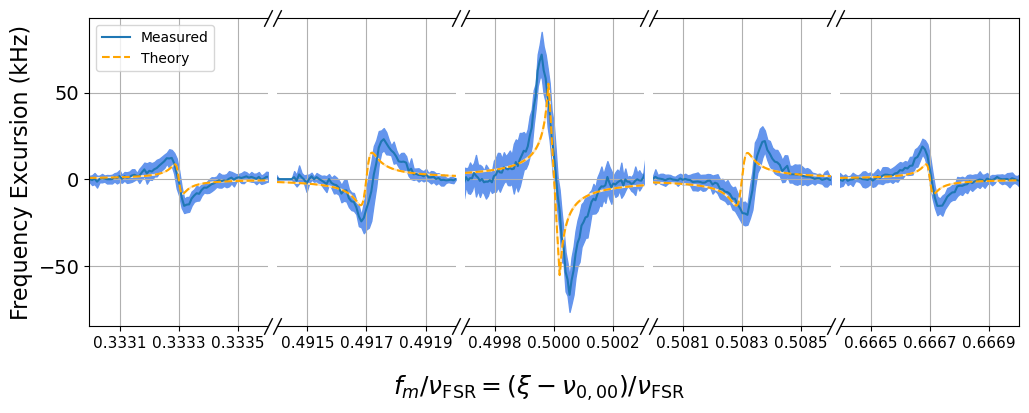}
\par\end{centering}
\caption{\protect\label{fig:FreqExcPlot} Plot of the measured and predicted frequency deviations of a PDH offset locked laser due to spurious optical frequencies interacting with various modes of the cavity. These are the five most prominent features in the center of Fig. \ref{fig:lock-shift-calc-plot}(a). We did not synchronously record the MZI balanced photodiode output with the frequency sweep, thus there is an ambiguity in the sign of the frequency deviations extracted from the trace data. We selectively inverted the extracted frequency deviations prior to averaging to obtain the plots shown here. The dark blue line is the averaged extracted frequency excursion from the data, the shaded light blue region is the corresponding standard deviation of the measured excursions at each sample point. The orange curve depicts theoretical predictions of the induced frequency excursion for a DSB modulation scheme. Error in the theory curve is too small to see in the plot above.}
\end{figure*}

A straightforward approach to measuring the excursions with this setup would be to sweep the offset frequency over one cavity FSR and look for phase excursions on the oscilloscope output. However, poor passive stability of the MZI path length difference precluded such a broad frequency sweep. We instead measured select regions of interest (ROI) from Fig. \ref{fig:lock-shift-calc-plot}(a) where we expect prominent spurious features. The result is Fig. \ref{fig:FreqExcPlot} which shows the residual for sweeps over the five most prominent features in the center of Fig. \ref{fig:lock-shift-calc-plot}(a).

In more detail, for each ROI we swept $f_{\text{m}}$ in a triangle wave with frequency $200\text{ Hz}$ and peak-to-peak amplitude $2.5\text{ MHz}$. During the sweep, the MZI voltage was digitized at $100\text{ ksps}$ forming a record of length $120\text{k}$ samples. Each record was subdivided into periods of the triangle wave and grouped into sets of $\sim40$ consecutive periods. The periods of each set were then averaged together to produce a single waveform per set. Each waveform was then phase corrected and fit to a sinusoid. The fit parameters were then used to normalize the waveform. Taking the arcsine gives the phase ramp. Subtracting out the corresponding phase ramps derived from the fit parameters revealed spurious phase excursions. Portions of the digitized record where the MZI phase was unfavorable were rejected. For each waveform, the timestamps of the maximum and minimum phase excursion were used to orient and align each waveform so that they could then be averaged together to produce a single final estimate of phase excursion vs. time. The time axis was then mapped to modulation frequency using the triangle wave parameters and the phase excursion axis mapped to frequency excursion using $\Delta L$. Fig. \ref{fig:FreqExcPlot} shows the measured frequency deviations as a function of modulation frequency, as a fraction of the cavity FSR, for the five most prominent features in Fig. \ref{fig:lock-shift-calc-plot}(a) (excluding those at the edges of the plot). Smaller features could be discerned with longer averaging windows, but the features shown here sufficiently illustrate our technique and confirm our estimations.

\section{Discussion}
The degree of cavity mode matching varies between setups. We choose alignment contrast $C_{1}\sim15$ and $C_{2}\sim30$ based on our experimental experience across eight PDH setups spanning UV to telecom colors. Reported values in the literature range from $C_{1}\sim5$, $C_{2}\sim3.5$ \cite{shi_suppression_2018} to as high as $C_{2}\sim5000$ \cite{mueller_determination_2000}. Since the spurious shift  $\Delta\xi_{\mathrm{spur}}$ scales as $1/C_{k}$ (Eq. \ref{eq:lock-shift-due-to-fake}), the shift of specific modes for a given setup can be readily rescaled from these contrasts.

Residual amplitude modulation (RAM) also produces lock point shifts in PDH systems. Because of this similarity, it is useful to compare the magnitude of shifts induced by RAM with those from unwanted spectral features. RAM originates from birefringence in the electro-optic crystal and parasitic etalons \cite{zhang_reduction_2014,shen_systematic_2015}, mechanisms distinct from those considered in this work. The RAM modulation depth has been experimentally measured \cite{zhang_reduction_2014,bi_suppressing_2019} and theoretically modeled \cite{shen_systematic_2015} to be on the order of $10^{4}-10^{5}\text{ \text{ppm}}$ with a standard EOM, corresponding to a lock point shift of $0.01-0.1$ times of the cavity linewidth. RAM has been reduced by more than $50\text{ dB}$ using active control \cite{zhang_reduction_2014} and $30\text{ dB}$ using wedge angle EO crystal \cite{bi_suppressing_2019}. At certain resonances, the shift $\Delta\xi_{\mathrm{spur}}$ (Fig. \ref{fig:lock-shift-calc-plot}) can exceed those from RAM. In the worst case, the total lasing frequency shift is the sum of the RAM-induced shift and $\Delta \xi_\text{spur}$.

The frequency shift in our model is static. In practice, however, it becomes time-varying when alignment contrasts $C_{k}$ fluctuate due to temperature drifts or mechanical vibrations, or when unwanted features power $P^{\prime}$ varies with EOM or RF amplifier temperature. Although we did not measure the noise spectrum of $\Delta\xi_{\mathrm{spur}}$, we expect similar behavior to RAM noise, which also arises from thermal and vibrational fluctuations. Indeed, RAM noise has been observed above the shot-noise floor up to $10-100\text{ Hz}$ Fourier frequency \cite{zhang_reduction_2014,li_analysis_2014}, much slower than typical PDH feedback loop bandwidth \cite{wang_practical_2025}.

\section*{Conclusion}
We compared serrodyne modulation with ordinary dual-sideband modulation in a practical context: a PDH offset lock that is continuously swept with realistic laser-cavity misalignment. For DSB modulation, the $-1$ spur can give rise to unwanted laser frequency shifts of up to $\sim\delta\nu_{c}/2$. For serrodyne modulation, where all spurs lie below $-13\text{ dB}$, the maximum unwanted shift is $\sim\delta\nu_{c}/40$. We measured the magnitude and location of a subset of these frequency shifts and found them to be in agreement with our model. To our knowledge this is the first systematic analysis and measurement of PDH lasing frequency deviation due to collision between unwanted optical cavity modes and optical spurs. We believe this analysis will help practitioners avoid foreseeable PDH locking errors and emphasizes the value of spectrally-pure optical offsets \cite{hildebrand_spectrally-pure_2024} for making PDH offset locks robust to laser misalignment.

\section{Back matter}
\textbf{Funding:} The authors acknowledge funding from the Army Research Laboratory (ARL) (Sponsor Award Number W911NF2420107: improved light-matter interfaces for state-control and readout of quantum systems).\\

\noindent\textbf{Acknowledgements:} Products or companies named here are included in the interest of completeness and does not imply endorsement by the authors or by the US government.\\

\noindent\textbf{Disclosures:} The authors declare no conflicts of interest.\\

\noindent\textbf{Data Availability:} The data that support the findings of this study are available from Roame Hildebrand upon reasonable request.

\section{References}

\begingroup
\renewcommand{\section}[2]{}

\bibliographystyle{bibtex/aip-modified}
\bibliography{bibtex/split-pdh_paper}

\end{document}